# Cross-Correlation Detection of Point Sources in WMAP First Year Data

J. Y. Nie[1], S. N. Zhang[1, 2]

[1] Department of Physics and Center for Astrophysics, Tsinghua University, Beijing 100084, P. R. China

[2] Key Laboratory of Particle Astrophysics, Institute of High Energy Physics, Chinese Academy of Sciences, Beijing 100049, P. R. China

## ABSTRACT

We apply a Cross-correlation (CC) method developed previously for detecting gamma-ray point sources to the WMAP first year data by using the Point-Spread Function of WMAP and obtain a full sky CC coefficient map. Analyzing this map, we find that the CC method is a powerful tool to examine the WMAP foreground residuals which can be further cleaned accordingly. Evident foreground signals are found in WMAP foreground cleaned maps and *Tegmark* cleaned map. In this process 101 point-sources are detected, and 26 of them are new sources besides the originally listed WMAP 208 sources. We estimate the flux of these new sources and verify them by another method. As a result, a revised mask file based on the WMAP first year data is produced by including these new sources.

## 1. Introduction

The Wilkinson Microwave Anisotropy Probe (WMAP)，as so far the most precise instrument for the full-sky anisotropy in the cosmic microwave background (CMB), has given the best measurements of several key parameters in cosmology (Spergel et al. 2003; Hinshaw et al. 2003; Bennett et al. 2003a). The results of WMAP have been applied in many fields of astronomy and brought some important influence (see, e.g., Tegmark et al. 2004; Nolta et al. 2004; etc.). A series of global maps of temperature fluctuations of CMB can be obtained from WMAP data products. From these maps some useful information can be derived, such as CMB temperature fluctuation power spectra, polarization and so on (Hinshaw et al. 2003; Kogut et al. 2003). These results are crucial for precision cosmology. Most parameters of modern cosmology can be determined by fitting the observed result of CMB with theoretic models (Peiris et al. 2003). Complete subtraction of foreground signals from the original full sky maps is important for reliable estimates of many cosmological parameters (Bennett et al. 2003b). On the other hand, WMAP is so far the only radio observatory covering the full sky in multi-band from 22.8 GHz (K band) to 93.5 GHz (W band). Searching for radio point sources is another application for these full-sky survey data, which is closely related to foreground subtraction.

In order to subtract foreground signals from WMAP data, the WMAP team has given a list of 208 radio point-sources (205 of them have been cross-identified in other band and the other 3 point sources have not be identified) (Bennett et al. 2003ab). The way to generate this list is to apply the filter $b_l/(b_l^2 C_l^{cmb} + C^{noise})$ on temperature maps which has been weighted by the square-root of the number of observations, where $b_l$ is the transfer function of the WMAP beam pattern, $C_l^{cmb}$ is the CMB angular power spectrum and $C_l^{noise}$ is the noise power. Then peaks that are greater than 5σ in the filtered maps are fit to a Gaussian profile plus a baseline plane. With this procedure the 208 point sources list is made.

In order to search for point sources not in the catalogue released by the WMAP team, we use a cross-correlation (CC) method by cross-correlating between CMB maps and WMAP Point-Spread Function (PSF). The calculated cross-correlation coefficients of pixels all over the sky make up a CC coefficient full-sky map which is then used to search for point sources. In the mean time, carrying some statistical analysis on this CC full sky map, we can also estimate the foreground residuals of radio point sources in a previously cleaned map, and eventually subtract the uncleaned foreground signals.

2. Cross-correlation

Cross-correlation is a common method in data analysis to compare the similarity between two data groups. We use CC method to analyze the CMB map with the Point-Source Function (PSF) of WMAP. The consistency between the PSF and the data in small local area in WMAP sky map can be estimated by this method. The consistency can be regarded as the criterion of point-source existence.

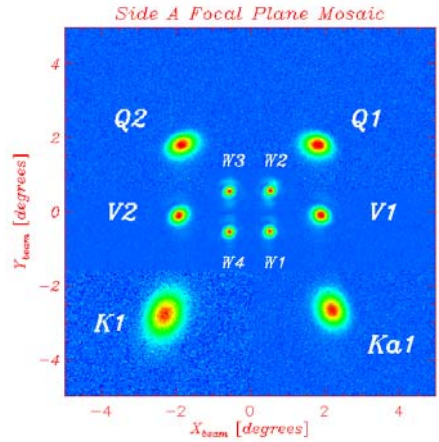

Fig.1 The beam profile of WMAP in different bands.

Because of the limited angular resolution of WMAP instruments, most radio sources can be considered as ideal point-sources. In the observed WMAP full sky maps, point sources are spread out as similar two-dimensional Gaussian peaks. In reality, the PSF of each WMAP instrument has its characteristic shape slightly different from an ideal symmetrical Gaussian function. But for the convenience of data processing, a two-dimensional Gaussian function is used as the PSF of each WMAP instrument.

We adopt the following equation,

$$C(n) = \frac{1}{m}\sum(D_i - \overline{D})(f_i - \overline{f}), \qquad (1)$$

to calculate the CC coefficients, where $C(n)$ is the value of CC coefficient at point $n$ (Hermsen 1983). This method was developed by the *COSB* team and was applied in *COSB* data analysis by Hermsen (1983) to search for gamma-ray point sources.

Theoretically, summation in Equation (1) should be carried out for all pixels in the sky.

Actually for pixels far away from the central location of a point source, the effect of the point source is negligible. Therefore we sum over only *m* pixels in a small circle of radius about five times of the full-width at half maximum (FWHM) of the PSF around the central point *n*. Therefore, $D_i$ is the value of pixel *i* in this small circular region, $i=1, 2......m$, and $\overline{D}$ is the average of $D_i$. $f_i$ is the value of PSF at pixel *i*, and $\overline{f}$ is the average of $f_i$, $i=1,2.....m$.

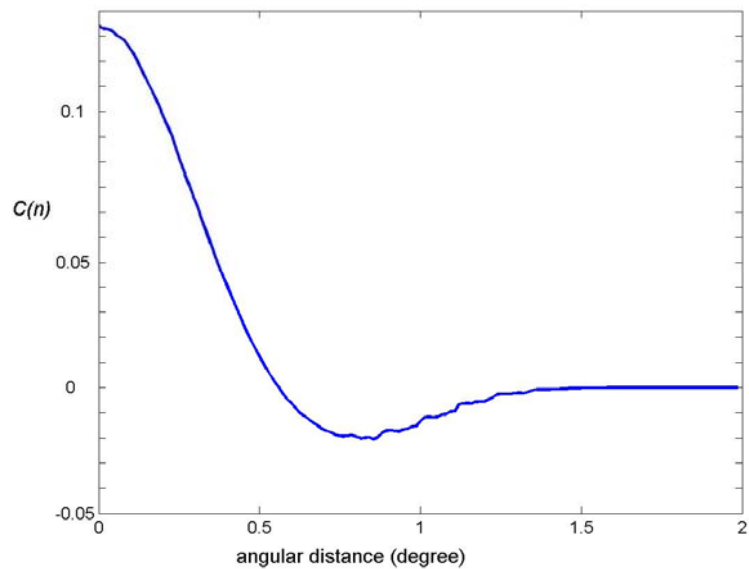

Fig.2 The distribution of CC coefficients around an ideal point source for the WMAP Q-band. The X-axis is for the angular distance and Y-axis is for the value of CC coefficient. Please note that negative CC values at about 0.8-1.1 degree. The CC value approaches to zero beyond about 1.5 degrees from the point source.

Fig.2 shows the CC coefficient distribution around an ideal point source. According to this figure the CC coefficient of a pixel has a local maximum value which is directly proportional to the source intensity, if the pixel is just at the location of a point source; the CC coefficients of pixels far away from point sources are zero. *C(n)* can thus be rewritten as

$$C(n) = \frac{A}{m} \sum_i (f_i - \overline{f})^2 , \qquad (2)$$

where *A* is directly proportional to the source intensity. Equation (1) can also be expressed in another form (all subscripts are omitted):

$$C(n) = \frac{1}{m^2} \left[ m \sum fD - (\sum f)(\sum D) \right] . \qquad (3)$$

Since the CMB temperature fluctuations follow Gaussian distribution, the CC coefficients *C(n)* also follow Gaussian distribution in the absence of point sources in the map. Therefore deviations of *C(n)* distribution away from Gaussian distribution are indication of uncleaned point sources in the map. It should also be noticed that at certain distances from the point source, the CC coefficients are negative; therefore uncleaned point sources can also produce deviations from Gaussian distribution for negative values of *C(n)*.

## 3. Dataset

We select three different data sources: the first one is the first year standard data products released by WMAP team, which are published on website of WMAP mission (http://lambda.gsfc.nasa.gov). Foreground cleaned maps, WMAP Internal Linear Combination Map（*ILC* Map）and *imap* data in five bands, which is without subtraction of foreground contamination, are adopted in our work. All these data are organized in the standard format of WAMP data products, which involves two one-dimensional arrays in every data file. These two arrays give the temperature and number of observations of every pixel over the full sky. The *HEALPix* scheme (http://healpix.jpl.nasa.gov/) is adopted to define the position and order of every element in arrays. In the *HEALPix* scheme, the whole sky is divided into $12 \times N^2$ small patches and every patch has a fixed serial number. For the data we selected, $N$ equals to 512, which means that the whole sky is divided into 3145728 small patches.

The second data source is the *Tegmark* cleaned map (Tegmark et al. 2003) which is considered to be cleaner than the standard WMAP foreground cleaned maps (Liu & Zhang 2005). The *Tegmark* cleaned map is also divided according to *HEALPix* scheme, but contains only one array for temperature fluctuations.

The third set of data comes from simulations. First, a theoretical power-spectrum is calculated from basic cosmological parameters published by the WMAP team. The second step is to simulate an ideal CMB temperature map from this theoretical power-spectrum. Third, smoothing is done using the published WMAP beam profiles. Finally, Gaussian noise of every pixel is added according to the number of observations of every pixel and instrument noise given by WMAP. By this way simulated CMB maps without any point source are produced.

## 4. Data processing

There are strong radio signal contaminations near the galactic equator in WMAP maps. This kind of strong foreground radio signal cannot be subtracted reasonably cleanly. Therefore those pixels near the galactic equator cannot be used. To remove those pixels, we applied the Kp2 mask file provided by WMAP, which is one of a series of mask files adopted in WMAP data analysis to subtract the pixels contaminated by galactic equator and strong point sources. The Kp2 mask removes galactic equator and several point sources; about 85% pixels from original 3145728 pixels are reserved. Considering our purpose of surveying for point sources, the Kp2 mask file is divided into galactic equator and point sources and we only use the first part to remove the pixels near the galactic equator. All of our analyses exclude pixels of the galactic equator.

A new full sky map made by CC coefficients can be produced using equation (1). To further reduce fluctuations due to Gaussian noise and CMB in small scale, a two-dimensional five-point Gaussian smoothing is carried over all pixels. Then possible point sources can be searched by finding local maxima in this map.

To verify the reliability of these point sources and estimate their fluxes, the temperature $T$ at the position of each point source is also calculated. For a point source, a series of nested concentric rings are made around this point source with width of 0.1 degree. The average temperature of these pixels within each ring is calculated as a function of radius. A Gaussian function is used to fit this function and derive the peak value as the temperature corresponding to this point source.

## 5. Point Source criterion

To identify a point source, we introduce the ratio CC/$\sigma_0$, where $\sigma_0$ is the parent standard deviation of CC coefficients in a field without any point source. The value of $\sigma_0$ is determined from two different datasets. The first one is *imap* data filtered by Kp2 mask. All CC coefficients are grouped according to the number of observations *N* at every pixel. For each individual group, the distribution of CC coefficients is fitted by a Gaussian distribution and $\sigma_0$ of different ranges of *N* are given. Then the relation between $\sigma_0$ and number of observations is fitted by an analytic function $\sigma_0 = \frac{a}{\sqrt{N}} + b$, as shown in Fig.3.

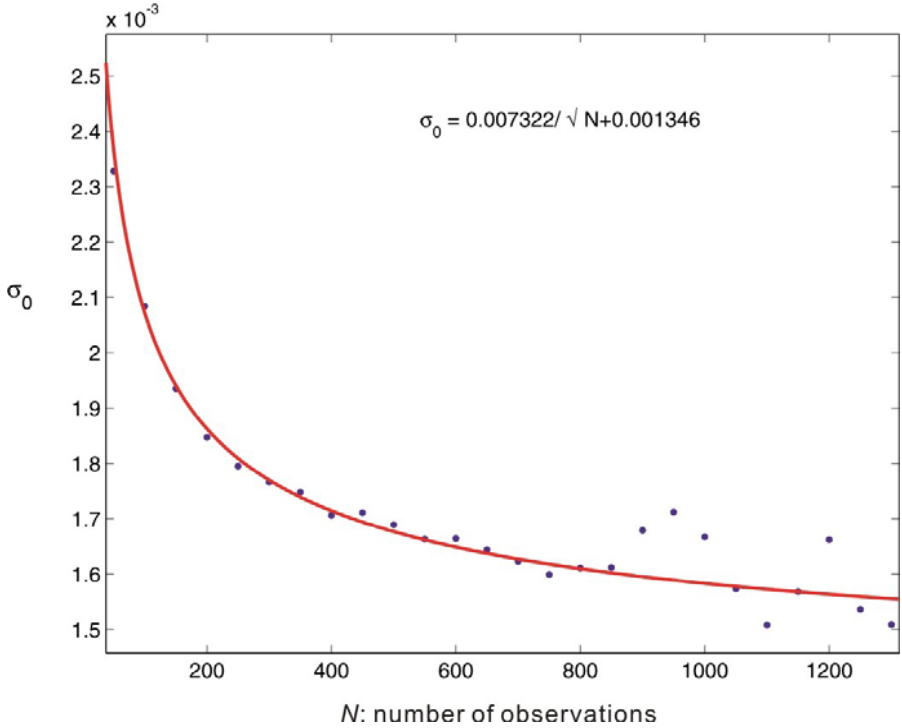

Fig.3 The relation between number of observations and $\sigma_0$.

Another dataset is the simulation data. Refer to the simulation procedure, the simulation data are without any signal of point source. So the mask file is not used. Using the same method, we get the coefficients *a* and *b* equal 0.0049 and 0.001 respectively. Therefore the difference of $\sigma_0$ from these two approaches are less than 5% and in our work we select the $\sigma_0$ which comes from true data. From the simulation data, we can also estimate the threshold value of CC/$\sigma_0$ for point source detection. In the simulation, we do not find any pixel with CC/$\sigma_0$ greater than 5. We therefore choose 5$\sigma_0$ as the criterion of point source detection.

6. Results
**6.1 Statistical distribution of CC coefficients**
Fig.4 (a) – (d) are the statistical distributions of CC coefficients of K, Q, V, W band WMAP uncleaned *imap* data and the corresponding simulation data (the X-axis is in units of $\sigma_0$). Panels (e) and (f) are results of WMAP *ILC* map and *Tegmark* cleaned map. From Fig.4 (a) – (d) we can find that the distributions of CC coefficients of the simulation data fit Gaussian distribution well. However the WMAP *imap* data only fit Gaussian distribution well around zero, and there are clear deviations greater than 3 $\sigma_0$ and less than -3 $\sigma_0$. Even for *Tegmark* cleaned data shown in Fig.3 (f) there are also many pixels with CC coefficients deviated from the Gaussian distribution. Point sources are the natural sources of these non-Gaussian fluctuations, because we have not masked out point sources in calculating these CC coefficients.

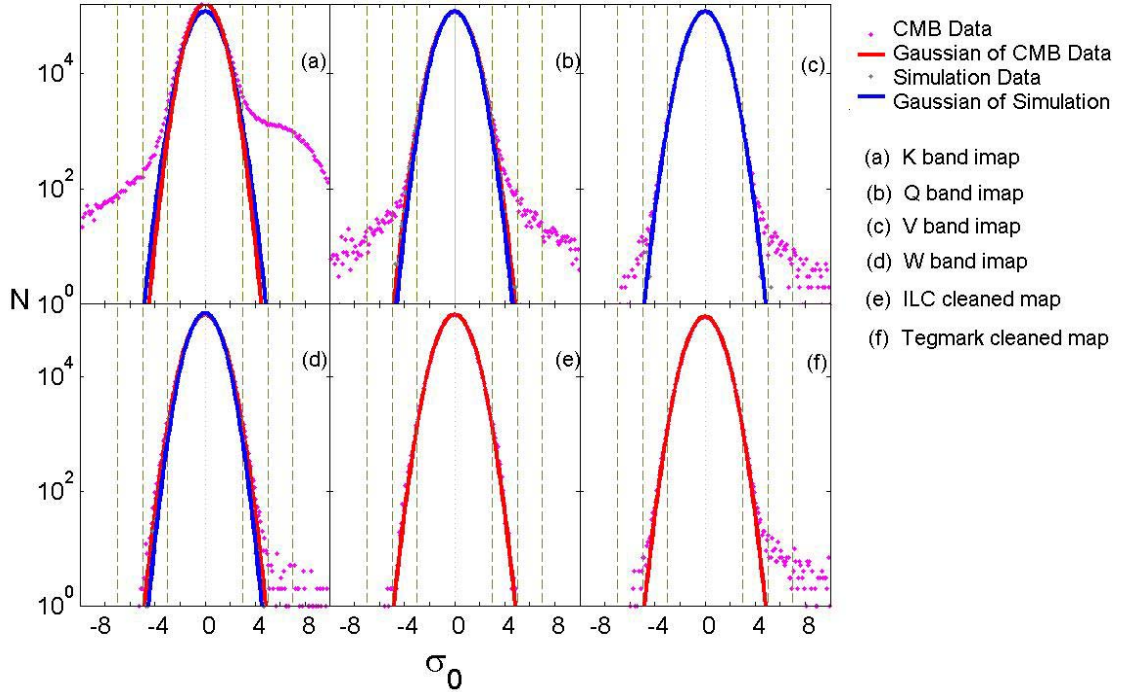

Fig.4 The statistics of CC coefficient. Purple points are true data and gray points come from simulation. Red and blue lines are the Gaussian fitting result for true and simulation data respectively. All fitted curves are scaled to the $\sigma_0$ and X-axis is the multiple relative to $\sigma_0$ of fitted result. Grey vertical dashed lines respectively correspond to $\pm 3, \pm 5, \pm 7$ $\sigma_0$.

According to simulation data, the CC coefficients of all pixels are within the range of -5 $\sigma_0$ to 5 $\sigma_0$. Because the simulation is the synthesis of the CMB fluctuation and Gaussian noise, we can draw a conclusion that the CMB and the noise cannot make the CC coefficients exceed $\pm 5$ $\sigma_0$. Thus those pixels with CC coefficients beyond $\pm 5$ $\sigma_0$ should be caused by some foreground contamination, which may be point sources. There are also clear excess of CC coefficient counts in the range of 3 $\sigma_0$ to 5 $\sigma_0$. The CC coefficients in this range may be caused by other unknown foreground residuals or noise; however we cannot distinguish between these two possibilities only from the CC coefficients of those pixels. Another possibility is that the excess of CC coefficient counts in the range 3 $\sigma_0$ to 5 $\sigma_0$ and the range beyond 5 $\sigma_0$ are caused by the same point sources, because several local maxima with different values and at different positions may be produced by one point source.

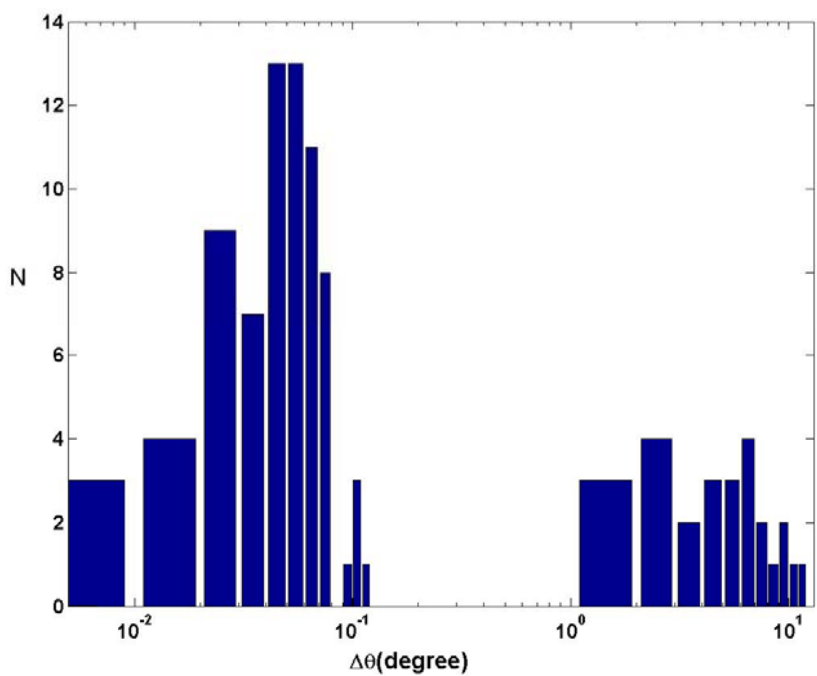

Fig.5 The histogram of distances for 101 sources detected by local maxima of CC coefficient to the nearest sources in WMAP 208 sources list. X-axis is the angular distance. Y-axis is the sources count in each distance interval. The 101 points are divided into two groups. The first group shown in the left include all distances less than 0.15 degree, therefore the sources detected by CC coefficient and corresponding nearest sources in 208 sources list can be regarded as the same sources. Distances in another group are all greater than 1 degree. We therefore regard them as new sources not in WMAP 208 sources list.

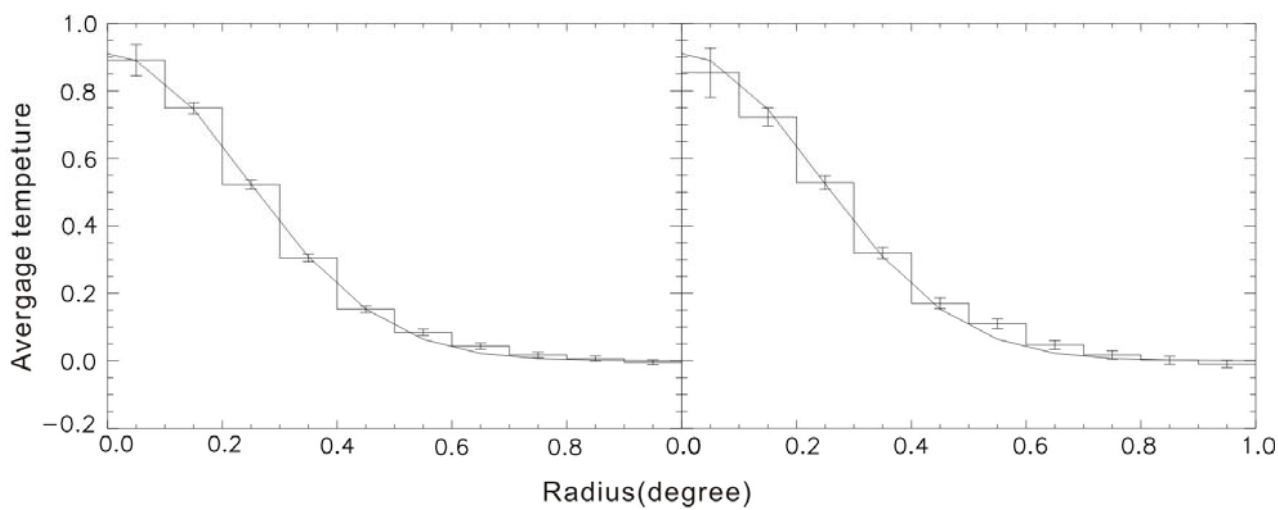

Fig.6 Consistency of the concentric-ring-averaging temperature and WMAP Q band PSF. In each panel the histogram is the average temperature in different radius from the central points, and the curve is WMAP PSF. The data of all point sources are normalized and then weight-averaged. The left panel is the average of 15 point sources masked by WMAP Kp2 mask, and the right panel is the average of 11 new point sources not included in Kp2 mask. The total $\chi^2$ values for the left and right panels are 14.2 and 18.0 for 9 degrees of freedom, respectively.

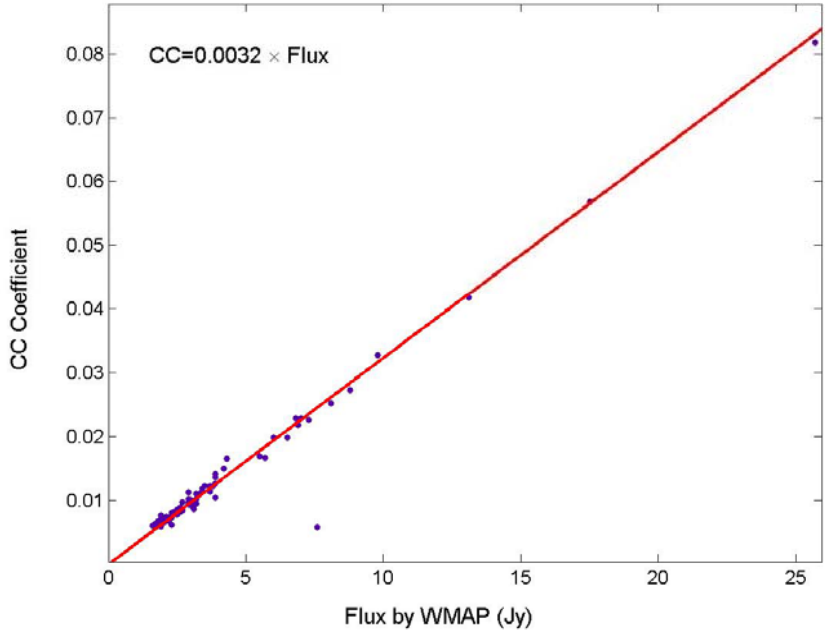

Fig.7 Linearity of the CC coefficient and Q band intensity given by WMAP, for these 75 sources in the first group shown in Fig.5.

**6.2 Point source detection**

Applying the local-maxima method to CC coefficients of Q band data, we find 101 local maxima from the pixels with CC coefficients beyond $5\sigma_0$. The comparison between the coordinates of these 101 points and the WMAP published 208 point sources is shown in Fig.5. Obviously these 101 points can be divided into two groups. One group has 75 points, which have their distances to the nearest point sources in WMAP list less than 0.15 degree. The distances of other 26 points in another group are greater than 1 degree. Taking into account of the angular resolution of WMAP in Q band, we regard that the 75 points in the first group as the corresponding sources in the WMAP 208 point sources. The second group of 26 points are thus considered as possible new point sources not included in the WMAP 208 sources.

We also compare the locations of our 26 new point sources with WMAP Kp2 mask files of both the first and three year data releases; 15 of our 26 new sources are included in both Kp2 masks. Other 11 point sources are missed in Kp2 mask. In Fig.6 we plot the average temperature at different radius around the location of these 26 points and find that the temperature distribution fits the WMAP Q band PSF well, indicating that these sources are consistent to point sources within the angular resolution of the WMAP Q band beam. To eliminate the contamination of these missed point sources, a new mask file including these 11 missed point sources is necessary.

As mentioned above, the CC coefficient $C(n)$ at the center of a point source is proportional to the intensity of this point source. For the 75 point sources in the first group, the CC coefficients and the Q band intensity determined by the WMAP team are correlated in Fig.7. As expected, a good linearity is present. The intensity of other 26 sources in the second group can be estimated by this linearity. The coordinates and Q band intensity of these 26 new point sources are listed in table 1.

.

Table 1    Position and intensity of 26 new sources.

|    | L[°]  | b[°]  | CC     | Flux[Jy] | Masked by Kp2 mask |
|----|-------|-------|--------|----------|--------------------|
| 1  | 302.0 | -45.0 | 0.0074 | 2.3      | N                  |
| 2  | 303.3 | -43.9 | 0.0068 | 2.1      | N                  |
| 3  | 277.3 | -36.1 | 0.0133 | 4.2      | Y                  |
| 4  | 280.8 | -35.5 | 0.0070 | 2.2      | N                  |
| 5  | 254.9 | -33.8 | 0.0065 | 2.0      | Y                  |
| 6  | 278.3 | -33.3 | 0.0060 | 1.9      | Y                  |
| 7  | 280.3 | -33.3 | 0.0060 | 1.9      | Y                  |
| 8  | 277.6 | -32.3 | 0.0062 | 1.9      | N                  |
| 9  | 281.9 | -32.1 | 0.0059 | 1.8      | N                  |
| 10 | 279.5 | -31.7 | 0.1117 | 34.9     | Y                  |
| 11 | 187.5 | -20.7 | 0.0079 | 2.5      | Y                  |
| 12 | 9.3   | -19.6 | 0.0391 | 12.2     | Y                  |
| 13 | 263.1 | -16.7 | 0.0063 | 2.0      | Y                  |
| 14 | 213.7 | -12.6 | 0.0250 | 7.8      | Y                  |
| 15 | 72.7  | -9.4  | 0.0067 | 2.1      | N                  |
| 16 | 75.5  | -8.8  | 0.0062 | 1.9      | N                  |
| 17 | 227.7 | -7.8  | 0.0067 | 2.1      | N                  |
| 18 | 224.4 | 2.5   | 0.0063 | 2.0      | N                  |
| 19 | 227.7 | 3.1   | 0.0112 | 3.5      | Y                  |
| 20 | 278.0 | 4.3   | 0.0066 | 2.1      | N                  |
| 21 | 318.2 | 10.9  | 0.0061 | 1.9      | Y                  |
| 22 | 327.6 | 14.6  | 0.0062 | 1.9      | N                  |
| 23 | 34.9  | 17.6  | 0.0121 | 3.8      | Y                  |
| 24 | 308.8 | 17.9  | 0.0104 | 3.3      | Y                  |
| 25 | 309.6 | 19.4  | 0.1197 | 37.4     | Y                  |
| 26 | 21.2  | 19.6  | 0.0103 | 3.2      | Y                  |

\* The error of the CC coefficient is about 0.0012, corresponding to the error of point source flux approximately equal to 0.4Jy

**6.3 Consistency check of these point sources**

Using the concentric-ring-averaging method described in section 4, the corresponding temperatures of all WMAP 208 sources are calculated and correlated with their intensity in Q band determined by the WMAP team, as shown in Fig.8; a good linearity is clearly seen (There are only 185 sources in this figure because intensities of other 23 sources were not given in WMAP source list). There is one point significantly far away from the fitted line, which is excluded in the linear fitting. The reason for this deviation will be explained later. Applying this linear function to the 101 point sources found with local maxima of CC coefficients, their intensities can be derived from their temperatures determined in the same way. Another set of intensity of these 101 point sources are also obtained from CC coefficient by the linearity shown in Fig.7. Fig.9 shows the linear correlation between these two sets of source intensities, indicating the consistency of two methods and thus reliability of the 101 point sources detected in this work.

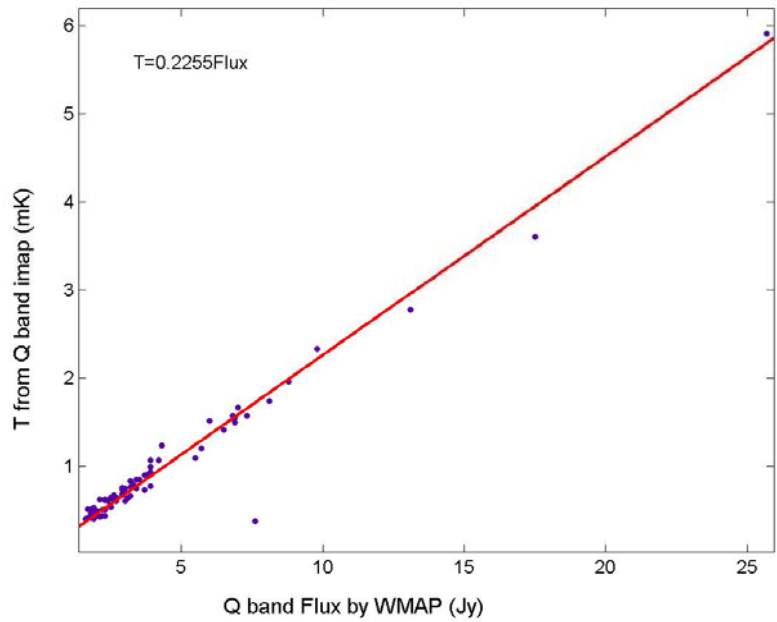

Fig.8 The linearity of correlation between Q band flux and temperature at the corresponding positions in the WMAP sky map.

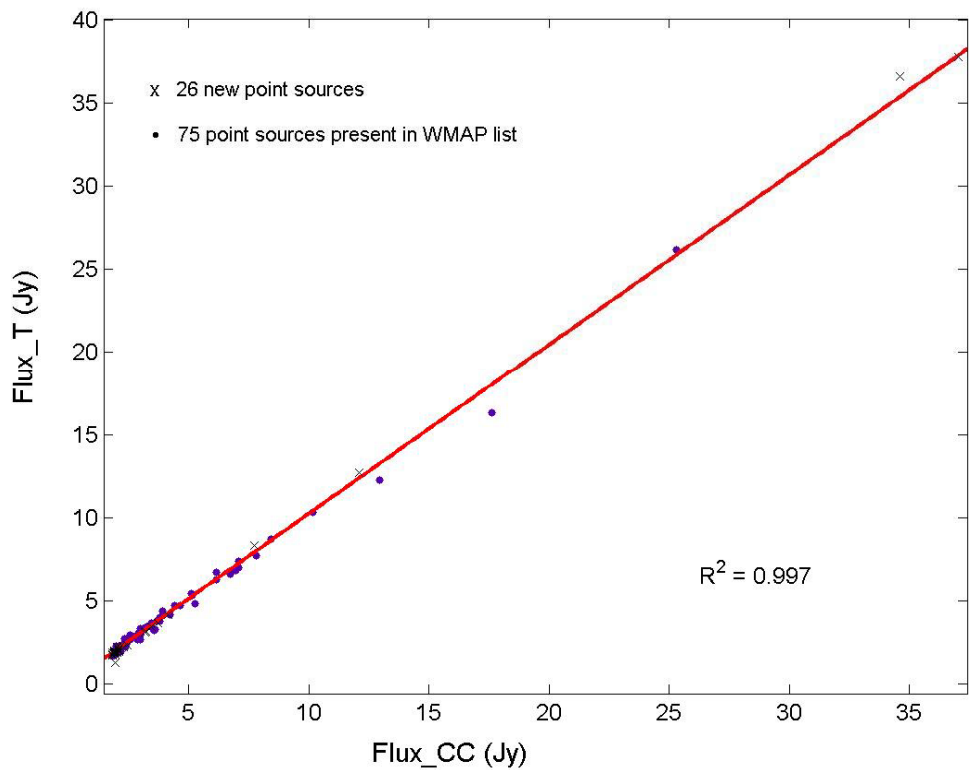

Fig.9 The linearity of correlation between the two sets of flux derived from two different ways for the 101 point sources. The X-axis is for flux derived from local maxima of CC coefficients and Y-axis is for those derived from temperatures at the corresponding positions in full sky map.

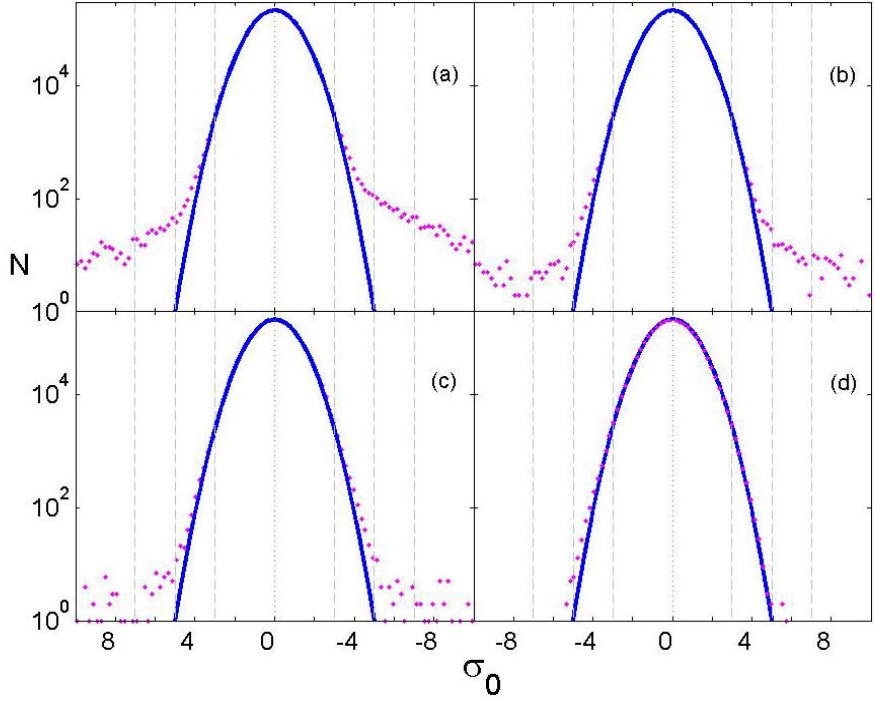

Fig.10 Statistics of CC coefficient using different foreground subtraction method. X-axis is the multiple of $\sigma_0$, gray vertical lines indicate $\pm 3$, $\pm 5$, $\pm 7\sigma_0$, respectively。 The purple dots are actual data. The blue lines are fitting result. There are different point source subtraction methods for Q band data. (a) is full foreground cleaned WMAP map （the corresponding *Tegmark* map result is in Fig.4 (f)） without any point sources subtraction. (b) subtract WMAP 208 sources, (c) subtract full Kp2 mask, (d) subtract WMAP Kp2 mask, WMAP 208 sources and our 26 new point sources.

From the linearity relation of CC coefficient and intensity of sources, with the $5\sigma_0$ detection threshold of CC coefficient, the corresponding intensity threshold is about 1.9 Jy. Comparing with the WMAP 208 source list, about 50 percent of sources with intensity 1.8-2.0 Jy in WMAP source list are detected by us; this agrees with theoretical expectation for detection sensitivity of an instrument (Zhang & Ramsden, 1990). For other 61 sources with intensities greater than 2 Jy in WMAP list, only two are not detected by our method.

For these two undetected sources with intensity greater than 2 Jy, their intensity values are 2.2±0.3, 2.4±0.4 Jy, respectively. It is reasonable that these sources cannot be detected because of the uncertainties of their intensity. It is noticeable that in Fig.8, there is one point obviously deviated from the fitting line; it is the point with intensity of 7.6±3 Jy in the WMAP list. According to the CC coefficient at the source location, the estimated intensity is only about 1.8 Jy. As a matter of fact the intensity of this source has been revised to 0 Jy in the new WMAP three year result, in good agreement with our result (0 means that the detection significance for this point source is weak, i.e., with intensity less than $2\sigma$ in this band).

The consistency of these two methods verifies the reliability of the CC method in point source detection. Although there is a detection threshold limitation which needs to be improved, applying this method to data of other WMAP bands should allow more point sources to be detected. A new catalog of point sources based on multi-band WMAP full sky survey is expected in our further work.

### 6.4 Revised mask file

The distribution of CC coefficients for an ideal CMB (with Gaussian noise) map must be a Gaussian distribution, and foreground contamination may cause deviations from Gaussian distribution. Therefore CC coefficient provides a powerful tool to check the effectiveness of the foreground subtraction. The WMAP foreground cleaned map (Fig.10(a)) and the *Tegmark* cleaned map (Fig.4(f)) are analyzed using CC coefficient, and found that both maps contain a lot of signals possibly from point sources. Fig.10 (b) shows the CC coefficient distribution after excluding the WMAP 208 point sources. There are still obvious deviations from Gaussian curve, suggesting the existence of more point sources. Many point-sources or point-like sources, which include 182 point sources in the WMAP 208 sources list and 15 sources of our 26 new point sources, are removed by using the Kp2 mask. Applying the full Kp2 mask on Q band cleaned map, Fig.10 (c) is obtained, which is much cleaner but still with some pixels exceeding $\pm 5\sigma_0$. Then *all* 26 new sources detected in this work and *all* WMAP 208 sources are subtracted from Fig.10(c) and Fig.10 (d) is produced; in this case the distribution of CC coefficients fits with Gaussian distribution well. This is the most cleaned map among these four and much less contaminated by point sources.

We have discussed that around a point source, negative CC coefficients such as that shown in Fig.4 (a)-(d) and Fig.10 (a)-(c), can also be produced, as shown in Fig.2. However after all point sources are subtracted from Q band *imap* data, convolving with the PSF for the determined intensities, we can find that most pixels with negative CC coefficient less than $-5\sigma_0$ have disappeared in Fig.10 (d).

According to the above results, a revised mask file is made. This new mask file includes the full WMAP Kp2 mask, 26 point sources in the WMAP 208 source list but missed in the Kp2 mask file, and also 11 new point sources (of the 26 new pointed sources detected in our work) not included in the Kp2 mask file. Using this revised mask file in data analysis, we can achieve a cleaner full sky map than the original Kp2 mask.

### 6.5 Comparison with the WMAP three year products

The WMAP team released their three year data products on Mar 16, 2006 (Hinshaw et al. 2006). In the three year data products, the point sources list is expanded to 373 sources (including first year 208 point sources). The expansion of WMAP source list is a result of higher Signal-to-Noise ratio owing to more number of observations. In the first year WMAP result, the threshold of detection of point sources is about 2.0 Jy in all bands. More observation time brings a lower threshold. So most additional point sources in WMAP new 373 sources list are weaker than 2.0 Jy. Only one new point source of our 26 new sources is included in WMAP newest point-sources list because most our new 26 sources are brighter than 1.9-2.0 Jy, which does not overlap the flux range of these additional point sources in WMAP three year list.

### 7. Summary and discussion

We have discussed some applications of CC method in WMAP data analysis, then verified the effectiveness and reliability of this method. The deviation between distribution of CC coefficients and Gaussian distribution implies the existence of point sources in the map. We have searched for local maxima of CC coefficients from pixels with CC coefficients over $5\sigma_0$ as candidate point sources from WMAP Q band data; a total of 101 sources are detected and 26 of them are new point sources, i.e., not detected previously in any WMAP band. Comparing with only 61 point sources detected by WMAP in Q band, our method is more sensitive in point source detection. The intensities of these point sources are also compared with results derived from other method, and confirmed the consistency between them. According to the result of simulation, the CC method could only detect point sources in pixels beyond $5\sigma_0$, corresponding to a flux density threshold about 1.9 Jy, due to the combined effects of CMB fluctuations and WMAP antenna noise.

CC coefficient can also be used to examine the foreground contamination of a sky map. Using CC coefficient, we found there are still some point sources remained in WMAP foreground cleaned maps and *Tegmark* cleaned map. A revised Kp2 mask file is made, by including the original WMAP 208 and our new 26 point sources. The distribution of CC coefficients fits the Gaussian curve well for the new map after removing those contaminated pixels using this new mask file.

Our above results indicate that the CC method is effective and accurate in point source detection in WMAP data. Our next step is to apply this method to WMAP data in all bands including the three year data. We anticipate more point sources to be detected and thus a cleaner foreground-free sky map is produced eventually.

Acknowledgement: The authors kindly thank Professor T. P. Li, for comments and suggestions, which helped us to improve the method of data processing. We are very grateful to the anonymous referee whose extremely detailed and insightful comments and suggestions allowed us to clarify several issues and improve the readability of this paper. We acknowledge the use of the LAMBDA. Support for LAMBDA is provided by the NASA Office of Space Science. This work has used the software package HEALPix (hierarchical, equal area and iso-latitude pixelization of the sphere, http://www.eso.org/science/healpix), developed by K.M. Goshen et al. This research has made use of the Astrophysical Integrated Research Environment (AIRE) which is operated by the Center for Astrophysics, Tsinghua University. SNZ acknowledges partial funding support by the Ministry of Education of China, Directional Research Project of the Chinese Academy of Sciences and by the National Natural Science Foundation of China under project no. 10521001.